\documentclass[useAMS,usenatbib]{mn2e}
\usepackage{graphicx}
 
\def\be{\begin{equation}} 
\def\ee{\end{equation}}

\def\HI{\hbox{H~$\scriptstyle\rm I\ $}}

\def\gsim{\lower.5ex\hbox{\gtsima}} 
\def\lsim{\lower.5ex\hbox{\ltsima}} \def\gtsima{$\; \buildrel > \over 
\sim \;$} \def\ltsima{$\; \buildrel < \over \sim \;$} \def\prosima{$\; 
\buildrel \propto \over \sim \;$} \def\gsim{\lower.5ex\hbox{\gtsima}} 
\def\lsim{\lower.5ex\hbox{\ltsima}} 
\def\simgt{\lower.5ex\hbox{\gtsima}} 
\def\simlt{\lower.5ex\hbox{\ltsima}} 
\def\simpr{\lower.5ex\hbox{\prosima}} \def\la{\lsim} \def\ga{\gsim}

\def\gtsima{$\; \buildrel > \over \sim \;$} 
\def\ltsima{$\; \buildrel < \over \sim \;$} 
\def\gsim{\lower.5ex\hbox{\gtsima}} 
\def\lsim{\lower.5ex\hbox{\ltsima}} 
\def\simgt{\lower.5ex\hbox{\gtsima}} 
\def\simlt{\lower.5ex\hbox{\ltsima}} 
\def\simpr{\lower.5ex\hbox{\prosima}} 
\def\la{\lsim} 
\def\ga{\gsim}

\def\E3{{\cal E}_{\rm g}^{III}}

\title[Cool Lyman Alpha Emitters]{The cool side of Lyman Alpha Emitters} 
\author[Dayal, Ferrara \& Saro]{Pratika Dayal$^{1}$\thanks{E-mail: 
dayal@sissa.it (PD)}, Andrea Ferrara$^{2}$ \& Alexandro Saro$^{3,4}$  \\ 
$^{{1}}$ SISSA/International School for Advanced Studies, Via Beirut 2-4 Trieste, Italy, 34014\\ 
$^{2}$ Scuola Normale Superiore, Piazza dei Cavalieri 7, 56126 Pisa, Italy\\
$^{3}$ Dipartimento di Astronomia dell'Universita di Trieste, via Tiepolo 11, I-34131 Trieste, Italy\\
$^{4}$ INFN, National Institute for Nuclear Physics, Trieste, Italy }

\begin{document} 
 
\date{Received 2008 December 24; in original form 2008 December 24} 
 
\pagerange{\pageref{firstpage}--\pageref{lastpage}} \pubyear{2002} 
 
\maketitle 
 
\label{firstpage} 
\begin{abstract} 
We extend a previous study of Lyman Alpha Emitters (LAEs) based on hydrodynamical cosmological simulations, by including two physical processes important for LAEs: (a) Ly$\alpha$ and continuum luminosities produced by cooling of collisionally excited \HI in the galaxy, (b) dust formation and evolution; we follow these processes on a galaxy-by-galaxy basis. \HI cooling on average contributes 16-18\% of the Ly$\alpha$ radiation produced by stars, but this value can be much higher in low mass LAEs and further increased if the \HI is clumpy. The continuum luminosity is instead almost completely dominated by stellar sources. The dust content of galaxies scales with their stellar mass, $M_{dust} \propto M_*^{0.7}$ and stellar metallicity, $Z_*$, such that $M_{dust} \propto Z_*^{1.7}$. As a result, the massive galaxies have Ly$\alpha$ escape fraction as low as $f_\alpha=0.1$, with a LAE-averaged value decreasing with redshift: $\langle f_\alpha \rangle =(0.33,0.23)$ at $z =(5.7,6.6)$. The UV continuum escape fraction shows the opposite trend with $z$, possibly resulting from clumpiness evolution. The model successfully reproduces the observed Ly$\alpha$ and UV luminosity functions at different redshifts and the Ly$\alpha$ equivalent width scatter to a large degree, although the observed distribution appears to be more more extended than the predicted one. We discuss possible reasons for such tension.
\end{abstract}


\section{Introduction}
Lyman Alpha Emitters (LAEs) are galaxies identified by means of their Ly$\alpha$ emission line, the strength and width of which makes the detection unambiguous to a large extent. Recently, there has been a massive surge in the amount of data available on these high redshift galaxies, specially those detected using the narrow band technique (Kashikawa et al. 2006; Shimasaku et al. 2006; Dawson et al. 2007; Gronwall et al. 2007; Murayama et al. 2007; Ouchi et al. 2008). This has enabled the Ly$\alpha$ and UV luminosity functions (LF) and the equivalent width (EW) distribution to be studied and understood statistically which has made LAEs important as both probes of reionization and high redshift galaxy evolution.

A number of theoretical models, both semi-analytic (Santos 2004; Dijkstra et al. 2007a,b; Kobayashi et al. 2007, 2009; Mao et al. 2007; Dayal et al. 2008) and those involving simulations (McQuinn et al. 2007; Nagamine et al. 2008) have been put forward to reproduce the observed data sets and put constraints on the ionization state of the IGM. However, the physical properties of LAEs including the star formation rates (SFR), metallicity, age and initial mass function (IMF) have been the subject of much debate.

To this end, in Dayal et al. (2009a) we used state of the art cosmological SPH simulations coupled with a Ly$\alpha$/continuum production/transmission model to infer the properties of LAEs. Using the SFR, age, metallicity for each galaxy as obtained from the simulation, a Salpeter IMF and the Early Reionization Model (ERM, Gallerani et al. 2007), we calculated the observed Ly$\alpha$ luminosity for each galaxy to identify the population that would be observed as LAEs based on the current observational limits. We found that at $z \sim 5.7$, the SFR are in the range $2.5-120\, {\rm M_\odot}$, the ages range from $38-326$ Myr, the average stellar metallicity, $Z_* \sim 0.22\, {\rm Z_\odot}$ and the color excess, $E(B-V)=0.15$. We thus showed that LAEs are intermediate age objects, with stellar metallicity higher than the generally assumed value of $5$\% ${\rm Z_\odot}$ and dust extinction that is very much consistent with other works (Lai et al. 2007; Nagamine et al. 2008). The only free parameters left in this work were the escape fractions of Ly$\alpha$ ($f_\alpha$) and continuum photons ($f_c$) from the galaxy, which we obtained by matching the model to the observations at each redshift.

This work extends and improves such previous study in many ways. The major novelty is 
represented by the inclusion of two important features: (a) a detailed model to calculate dust mass evolution and the corresponding optical depth to continuum photons enabling us to link 
$f_c$ to the physical properties of galaxies such as the SFR, IMF, age and gas mass; 
(b) the Ly$\alpha$ and continuum luminosities produced by cooling of collisionally 
excited \HI in the galaxy ISM.

The latter point is very interesting since most of the studies on LAEs have focused on stars as the main source of both Ly$\alpha$ and continuum luminosity; an exception is   Dijkstra (2009), in which the case of Ly$\alpha$ and continuum luminosity coming from cooling of collisionally excited \HI in the ISM alone has been explored. Cooling of \HI becomes even more important when one considers that Ly$\alpha$ blobs (LABs) are powered either by an inflow of cooling gas (Dijkstra \& Loeb 2009) or by supernovae (Mori, Umemura \& Ferrara 2004; Mori \& Umemura 2007). 

An additional motivation of this paper is to understand how LAEs become progressively more dusty and how their dust component affects their visibility along with their observed Ly$\alpha$ and UV continuum LFs, and line equivalent widths; in addition, gas emission might also have important effects on these quantities which need to be determined. The study is based on high resolution numerical simulations which enable us to study these effects accurately and on a galaxy-by-galaxy basis.

\section{Simulations}
\label{simu}
We use state-of-the-art cosmological SPH simulations, the details of which can be found in Dayal et al. (2009a). In brief, simulations of a cosmic size ($75h^{-1}$ comoving Mpc), have been carried out using a TreePM-SPH code GADGET-2 (Springel 2005) with the implementation of chemodynamics as described in Tornatore et al. (2007). The run assumes a metallicity-dependent radiative cooling (Sutherland \& Dopita 1993) and a uniform redshift-dependent Ultra Violet Background (UVB) produced by quasars and galaxies as given by Haardt \& Madau (1996). The code also includes an effective model to describe star formation from a multi-phase interstellar medium (ISM) and a prescription for galactic winds triggered by supernova (SN) explosions, Springel \& Hernquist (2003). In their model, star formation occurs due to collapse of condensed clouds embedded in an ambient hot gas. Stars with mass larger then 8${\rm M_\odot}$ explode as supernovae and inject energy back into the ISM. These interlinked processes of star formation, cloud evaporation due to supernovae and cloud growth caused by cooling lead to self-regulated star formation. The relative number of stars of different mass is computed for this simulation by assuming the Salpeter (1955) IMF between 1 and 100${\rm M_\odot}$. Metals are produced by SNII, SNIa and intermediate and low-mass stars in the asymptotic giant branch (AGB). Metals and energy are released by stars of different masses by properly accounting for mass--dependent lifetimes as proposed by Padovani \& Matteucci (1993).  We adopt the metallicity--dependent stellar yields from Woosley \& Weaver (1995) 
 and the yields for AGB and SNIa from van den Hoek \& Groenewegen (1997).  Galaxies are recognized as gravitationally bound groups of star particles by running a standard friends-of-friends (FOF) algorithm, decomposing each FOF group into a set of disjoint substructures and identifying these by the SUBFIND algorithm (Springel et al. 2001). After performing a gravitational unbinding procedure, only sub-halos with at least 20 bound particles are considered to be genuine structures, Saro et al. (2006). For each ``bona-fide'' galaxy in the simulation snapshot, we compute the mass-weighted age,
the total halo/stellar/gas mass, the SFR, the mass weighted gas/stellar metallicity, the mass-weighted gas temperature and the half mass radius of the dark matter halo.

\begin{figure} 
\center{\includegraphics[scale=0.45]{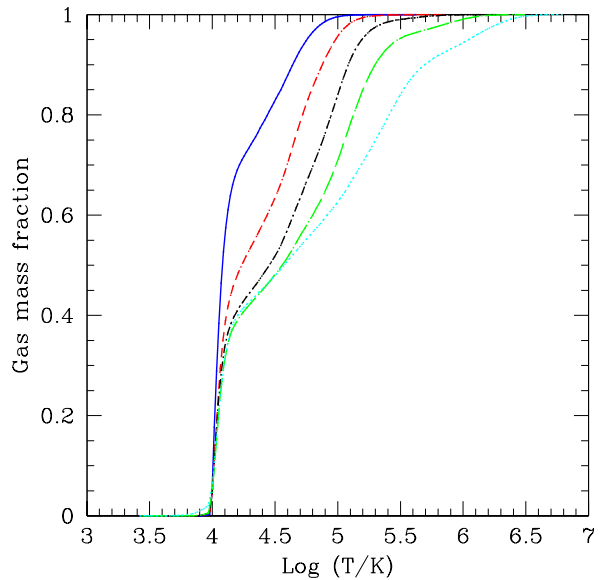}} 
\caption{Cumulative temperature distribution for interstellar gas averaged over all galaxies having dark matter halo masses in the range (in solar mass): $<10^{10}$ (solid line), $10^{10-10.4}$ (short-dashed), $10^{10.4-10.8}$ (dash-dotted), $10^{10.8-11.2}$ (long-dashed), $>10^{11.2}$ (dotted).}
\label{temp} 
\end{figure}

We compute the intrinsic and observed values of the Ly$\alpha$/continuum luminosity for all the structures identified as galaxies in the simulation boxes at the redshifts of interest ($z\sim 5.7, 6.6$). Of these, galaxies with an observed Ly$\alpha$ luminosity, $L_\alpha \geq 10^{42.2}\, {\rm erg\, s^{-1}}$ and observed equivalent width, $EW \geq 20$ \AA\, \footnote{Since we assume a simple scaling to obtain the escape fraction of Ly$\alpha$ relative to continuum photons, all galaxies are classified as LAEs based on the EW criterion.} are identified as LAEs, following the current observational criterion for LAE identification.

The adopted cosmological model for this work corresponds to the $\Lambda$CDM Universe with $\Omega_{\rm m }=0.26, \Omega_{\Lambda}=0.74,\ \Omega_{\rm b}=0.0413$, $n_s=0.95$, $H_0 = 73$ km s$^{-1}$ Mpc$^{-1}$ and $\sigma_8=0.8$, thus consistent with the 5-year analysis of the WMAP data (Komatsu et al. 2009).

\section{Intrinsic luminosities}
\label{lyag}

As mentioned before, the huge amount of literature on LAEs has generally been focused on stars as the main source of both Ly$\alpha$ and continuum luminosity. However, by virtue of using our SPH simulation, we can consistently calculate the separate stellar and cooling \HI gas contributions to the total Ly$\alpha$ and continuum luminosities, as discussed in the following subsections.

\begin{figure*} 
\center{\includegraphics[scale=1.0]{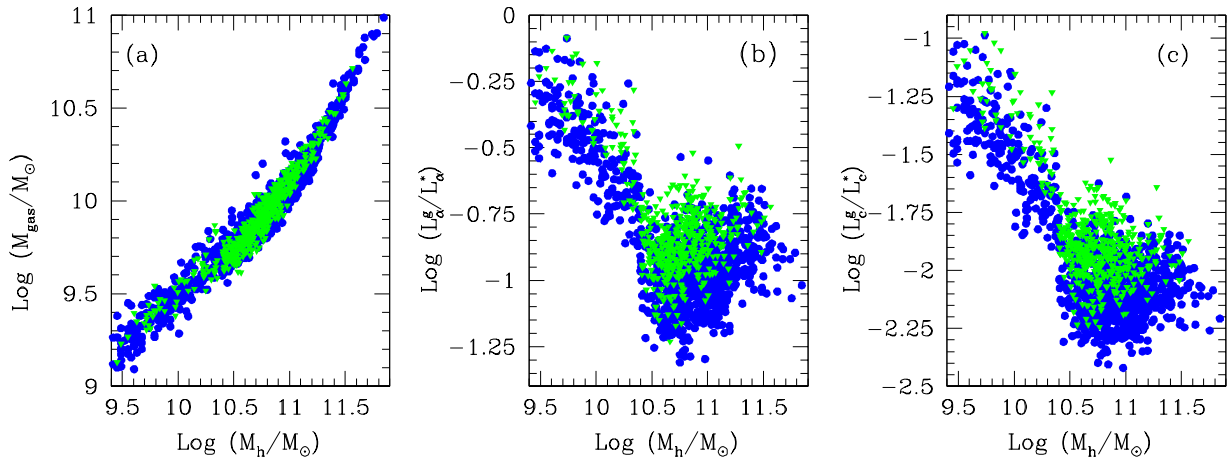}} 
\caption{(a) The total gas mass $M_{gas}$; (b) ratio of cooling Ly$\alpha$ luminosity from collisionally excited \HI gas, $L_\alpha^g$, to that from stars, $L_\alpha^*$; (c) same ratio for the continuum luminosity at 1375 \AA, for z=5.7 (circles) and z=6.6 (triangles) shown as a function of the halo mass, $M_h$, for the galaxies identified as LAEs at these redshifts.}
\label{rat} 
\end{figure*} 

\subsection{Star powered luminosity}
\label{sp_lum}
Star formation produces photons with energies $> 1$~Ryd, which ionize the \HI in the ISM. Due to the high density of the ISM, recombinations take place on a short time scale and this produces Ly$\alpha$ luminosity. We represent this luminosity as $L_\alpha^*$, or as the Ly$\alpha$ produced due to stars, which is calculated as
\begin{equation}
L_\alpha^* = \frac{2}{3} Q (1-f_{esc}) h \nu_\alpha,
\end{equation}
where $Q$ is the production rate of \HI ionizing photons, $f_{esc}$ is the fraction of these ionizing photons which escape the galaxy without causing any ionizations, $h$ is the Planck constant and $\nu_\alpha$ is the frequency of the Ly$\alpha$ photons. The factor of two-thirds arises assuming hydrogen Case B recombination (Osterbrock 1989) and we use $f_{esc} \sim 0.02$, following the results of Gnedin et al. (2008).

The continuum luminosity ($L_c^*$)\footnote{The continuum luminosity is measured in the wavelength range between 1250-1500\AA, centered at 1375 \AA.} associated with star formation and $Q$ are both obtained from {\rm STARBURST99} (Leitherer et al. 1999), using the stellar metallicity ($Z$), age ($t_*$), IMF and the SFR ($\dot M_*$) for each galaxy as obtained from the simulation. The use of the constant SFR is justified by the fact that all the galaxies identified as LAEs, at both the redshifts considered, have mass-weighted ages between 30-300 Myr, which implies that recent star formation does not contribute significantly to the star powered Ly$\alpha$ and continuum luminosity.


As a quantitative illustration, for a galaxy with $t_*=200 \, {\rm Myr}$, $Z = 0.2\, {\rm Z_\odot}$, $\dot M_* = 1\, {\rm M_\odot} \, {\rm yr^{-1}}$, we find $Q = 10^{53.47} {\rm s^{-1}}$ and the corresponding stellar Ly$\alpha$ luminosity for $f_{esc}=0.02$ is $L_\alpha^*=3.18 \times 10^{42} {\rm erg \, s^{-1}}$. For the same galaxy, the stellar continuum luminosity is $L_c^* = 3.5 \times 10^{40} {\rm erg \, s^{-1}\AA^{-1}}$, which yields a stellar EW of about $91$ \AA. Complete details of this calculation can be found in Dayal et al. (2008, 2009a).

\subsection{Luminosity from cooling \HI}
In addition to the formation of stars, our numerical simulations allow us to precisely track the evolution of the amount of gas present in each galaxy of the studied cosmic volume along with its temperature distribution. The cooling of collisionally excited \HI in the ISM of LAEs gives rise to both Ly$\alpha$ and UV continuum luminosities that add to those powered by stars. 
The Ly$\alpha$ luminosity, $L_\alpha^g$ [erg ${\rm s^{-1}}$], produced by the cooling of neutral hydrogen gas of total mass $M_{HI}$,  depends both on the number density of electrons and the number density of \HI. We assume the hydrogen gas in each galaxy to be concentrated within a radius $r_g$, such that $r_g = 4.5 \lambda r_{200}$, where the spin parameter, $\lambda=0.04$ (Ferrara, Pettini \& Shchekinov 2000) and $r_{200}$ is calculated assuming the collapsed region has an overdensity of 200 times the critical density at the redshift considered. Then, the density of electrons can be expressed as $(1-\chi_{HI}) M_{HI}[\mu m_H 4\pi r_g^3]^{-1}$ and the number density of \HI can be expressed as  $\chi_{HI} M_{HI}[\mu m_H 4 \pi r_g^3]^{-1}$, where $\mu$ is the mean molecular weight and $m_H$ is the proton mass. Then using the coefficients of $7.3\times 10^{-19} e^{-118400/T} \chi_{HI}\, [{\rm erg\, cm^3 \, s^{-1}}]$ as given by Dijkstra (2009), the total Ly$\alpha$ luminosity from cooling of collisionally excited \HI from the entire volume in which gas is distributed can be expressed as
\begin{equation}
L_\alpha^g = 7.3\times 10^{-19} \frac{ M_{HI}^2}{4\pi r_g^3 (\mu m_H)^2} G_\alpha(T,\chi_{HI}),   
\end{equation}
where 
\begin{equation}
G_\alpha(T,\chi_{HI})=\sum_{i=1}^N\chi_{HI, i}^2 (1-\chi_{HI, i})e^{-118400/T_i} f_{I,i}^2
\end{equation}
is a nondimensional function depending on the \HI temperature, $T$, and the fraction of neutral hydrogen, $\chi_{HI}$. Further, $f_I$ is the \HI mass fraction mass found in the $i$-th bin of the simulated temperature distribution, which is discretized into $N=100$ bins equally spaced in logarithm between $10^{3.4}$~K and $10^{6.8}$~K, to provide a smooth representation of $G_\alpha$. The value of the \HI fraction ($\chi_{HI,i}$) in each bin is calculated by inserting the temperature ($T_i$) of that bin into the collisional ionization-recombination rate equation for \HI (Cen 1992). Each galaxy in the simulation box is first assigned to one of the five halo mass ranges shown in Tab. \ref{temp_dist}. Then we add its gas mass to the corresponding temperature bin and weigh over the total mass in that mass range. The values of $G_\alpha(T,\chi_{HI})$ and galaxy-averaged temperature distributions are shown in Tab. \ref{temp_dist} and Fig. \ref{temp}, respectively. From the Figure, it appears that massive galaxies tend to have a larger fraction of their interstellar gas in a hot component. In eq. \ref{rat}, $M_{HI}$ is the total \HI mass of the galaxy assuming a primordial gas composition (76\% H, 24\% He).
\begin{table} 
\begin{center} 
\caption {Values of the functions $G_\alpha(T,\chi_{HI})$, $G_c(T)$ and the corresponding intrinsic EW from cooling of collisionally excited \HI gas in the ISM for different halo mass ranges.}
\begin{tabular}{|c|c|c|c} 
\hline 
$M_h$ & $G_\alpha(T,\chi_{HI})$ & $G_c(T)$ & $EW_i^g$ \\
$[M_\odot]$ & $$ & $$ & $[{\rm \AA}]$  \\ 
\hline
$ <10^{10}$ & $1.31 \times 10^{-7}$ & $0.50$ & $887.1$ \\
$10^{10-10.4}$ & $1.14 \times 10^{-7}$  & $0.45$ & $985.6$\\ 
$10^{10.4-10.8}$ & $7.32 \times 10^{-8}$ & $0.40$ & $1108.8$\\ 
$10^{10.8-11.2}$ & $9.18 \times 10^{-8}$ & $0.37$ & $1198.8$\\ 
$>10^{11.2}$ & $1.18 \times 10^{-7}$ & $0.34$ & $1304.5$\\
\hline
\label{temp_dist} 
\end{tabular} 
\end{center}
\end{table} 

Having determined the value of $L_\alpha^g$, the continuum luminosity, $L_c^g$ $[{\rm erg \, s^{-1}}]$, produced by cooling of collisionally excited \HI is calculated as 
\begin{equation}
L_c^g  = L_\alpha^{g}\, \phi(\nu_c/\nu_\alpha) \bigg(\frac{\nu_c}{c }\bigg)  G_c(T),
\end{equation}
where we consider the $2\gamma$ ($2s-1s$) transition to be the only source of the continuum and
\begin{equation}
G_c(T)= \sum_{i=1}^N \frac{\Omega(1s,2s)}{\Omega(1s,2p)} \bigg \vert_i f_{I,i}.
\end{equation}
Here $\nu_c$, $\nu_\alpha$ are the frequencies corresponding to the continuum (1375 \AA) and Ly$\alpha$ wavelengths (1216 \AA) respectively, $c$ is the speed of light and $\phi(\nu/\nu_\alpha) d\nu/\nu_\alpha$ is the probability that a photon is emitted in the frequency range $\nu \pm d\nu/2$, Spitzer \& Greenstein (1951). Again, $f_I$ is the \HI mass fraction mass found in the $i$-th bin of the simulated temperature distribution, which is discretized into $N=100$ bins equally spaced in logarithm between $10^{3.4}$~K and $10^{6.8}$~K. We use $\phi(\nu_c/\nu_\alpha)=3.1$, for the continuum frequency corresponding to 1375 \AA. The collisional excitation rates, $\Omega(1s, 2s)$ and $\Omega(1s, 2p)$ are calculated for each temperature bin by using the results of Anderson et al. (2000). Again using the temperature distribution for all galaxies as shown in Fig. \ref{temp}, the mass-weighted value of this ratio varies with the halo mass range as shown in Tab. \ref{temp_dist}. However, as will be shown later, the continuum from the stars always dominates over $L_\alpha^g$, which makes our results insensitive to the exact values of the collisional excitation rates and continuum wavelength chosen \footnote{The intrinsic EW from cooling of collisionally gas alone decreases from 1449 to 681 \AA, as the continuum wavelength increases from 1250 to 1500 \AA. However, since the stellar continuum always dominates that from the gas, we show the results at a wavelength corresponding to 1375 \AA, for convenience.}.  Using these conversion factors, the calculated value of intrinsic EW of the Ly$\alpha$ line from gas alone has a value between 887-1304 \AA, depending on the halo mass, as shown in Tab. \ref{temp_dist}. 

Once that $L_\alpha$ and $L_c$ from both the stars and ISM gas component have been determined, the total intrinsic equivalent width, $EW^{int}$, can be calculated as
\begin{equation}
EW^{int} = \frac{L_\alpha^* + L_\alpha^g}{L_c^* + L_c^g}.
\end{equation}
The {\it observed} equivalent width, $EW$, is affected by dust in the ISM (modifying both the continuum and Ly$\alpha$ luminosity) and transmission through the IGM, which only damps the Ly$\alpha$ luminosity. The observed EW in the rest frame of the galaxy is calculated as
\begin{equation}
EW = EW^{int} \bigg(\frac{f_\alpha}{f_c}\bigg) T_\alpha,
\end{equation}
where $f_\alpha/f_c$ represents the differential effect of dust on Ly$\alpha$ and continuum photons; a fraction $T_\alpha$ of the Ly$\alpha$ luminosity emerging from the galaxy is transmitted through the IGM. The transmission through the IGM depends on a number of factors including the SFR, the escape fraction of \HI ionizing photons, the size of the Str\"omgren sphere built by the galaxy, the global value of the fraction of \HI and the extent to which sources are clustered. Complete details of the calculation of $T_\alpha$ can be found in Dayal et al. (2009a). The effect of dust on both the continuum and Ly$\alpha$ luminosity 
is discussed in Sec. \ref{dust model}.

We find that the contribution of $L_\alpha^g$ to the total intrinsic Ly$\alpha$ luminosity is not negligible; the average value of $L_\alpha^g / L_\alpha^* =(0.16,0.18)$ at $z=(5.7,6.6)$ respectively, as seen from Panel (b), Fig. \ref{rat}. The ratio $L_c^g/L_c^*$, on the other hand is about a factor ten smaller, i.e. (0.01,0.02) at $z=(5.7,6.6)$, respectively (Panel (c) of same Figure). Adding $L_\alpha^g$ and $L_c^g$, therefore, has the effect of boosting up the intrinsic EW. As shown in Dayal et al. (2009a), the average intrinsic EW produced by stars only at $z\sim 5.7 = 95$ \AA. However, as shown in Tab. \ref{tab1}, we find that adding $L_\alpha^g$ boosts the average intrinsic EW to $106$ \AA, i.e., by a factor of about 1.12 as expected.

It is interesting to see that even though the gas mass (Panel (a), Fig. \ref{rat}) is almost similar for a given halo mass at $z \sim 5.7$ and $6.6$, both $L_\alpha^g$ and $L_c^g$ are slightly larger at $z \sim 6.6$. This can be explained noting that the virial radius decreases with increasing redshift, which leads to a corresponding increase in the \HI density. The higher density leads to an increase in the collisional excitation rate and hence, the cooling rate, producing more $L_\alpha^g$ and $L_c^g$.

Further, as shown in Dayal et al. (2009a), the SFR rises steeply with increasing halo masses, which leads to $L_\alpha^*$ and $L_g^*$ becoming increasingly dominant as compared to  $L_\alpha^g, L_c^g$. Hence, the contribution of ISM gas towards both the Ly$\alpha$ and continuum luminosity decreases with increasing halo mass. The dip in $L_\alpha^g$ (and the corresponding dip in $L_c^g$) for halo masses between $10^{10.4}$ and $10^{10.8}\, M_\odot$ is caused by the lower value of $G_\alpha(T,\chi_{HI})$ for these masses, as seen from Tab. \ref{temp_dist}.

\section{Dust Model}
\label{dust model}

Dust is produced by supernovae and evolved stars in a galaxy. However, several authors (Todini \& Ferrara 2001; Dwek et al. 2007) have shown that the contribution of AGB stars becomes progressively less important and at some point negligible towards higher redshifts ($z \gsim 5.7$). This is because the typical evolutionary time-scale of these stars ($\geq 1$ Gyr) becomes longer than the age of the Universe above that redshift. However, under certain conditions thought to hold in quasars, in which extremely massive starbursts occur, the contribution of AGB can become important somewhat earlier (see Valiante et al. 2009). We therefore make the hypothesis that the dust seen in LAEs at $z \ge 5.7$ is produced solely by type II supernovae. We compute the amount of dust in each LAE by applying the evolutionary model described below by post-processing the simulation outputs.

As mentioned before, for each simulated galaxy the SFR, $\dot M_*$, the total stellar mass, $M_*$ and the mass-weighted stellar age, $t_*$ are available. We then use the average SFR over the SF history of the galaxy as $\langle \dot M_* \rangle = M_* / t_*$ to calculate the dust mass. This is because the SNII rate, and hence the dust content of the galaxy depend on the entire SFR history and not just the final values obtained from the simulation. The SNII rate is estimated to be $\gamma \langle \dot M_* \rangle$, where $\gamma \sim (54 \,{\rm M_\odot})^{-1}$ for a Salpeter IMF between the lower and upper mass  limits of $1$ and ${100\, \rm M_\odot}$ respectively (Ferrara, Pettini and Shchekinov 2000). We have assumed that the progenitors of SNII have a mass larger than ${8\, \rm M_\odot}$ and in order to simplify the calculation, we have adopted the instantaneous recycling approximation (IRA), i.e., the lifetime of stars with mass larger than $8\, {\rm M_\odot}$ is zero.

Using this SNII rate, we can calculate the evolution of the total dust mass, $M_{dust}(t)$, in the galaxy as
\begin{equation}
\label{md}
\frac{d M_{dust}(t)}{dt} = y_d \gamma \langle \dot M_* \rangle - \frac{M_{dust}(t)}{\tau_{dest}(t)} - \frac{M_{dust}(t)}{M_{g}(t)} \langle \dot M_* \rangle,
\end{equation}
where  the first, second and third terms on the RHS represent the rates of dust production, destruction and astration (assimilation of a homogeneous mixture of gas and dust into stars) respectively; the initial dust mass is assumed to be zero. Further, $y_d$ is the dust yield per SNII, $\tau_{dest}$ is the timescale of dust destruction due to SNII blast waves and $M_{g}(t)$ is the gas mass in the galaxy at time $t$. For $y_d$, the average dust mass produced per SNII, we adopt a value of ${0.5\, \rm M_\odot}$ (Todini \& Ferrara 2001; Nozawa et al. 2003, 2007; Bianchi \& Schneider 2007). The amount of gas left in the galaxy, $M_g(t)$, at any time $t$ is calculated as ${d M_g(t)}/{dt} = -\langle \dot M_* \rangle$ and each galaxy is assumed to have an initial gas mass given by the cosmological baryon to dark matter ratio $\Omega_b/\Omega_m$. The final gas mass so obtained for each galaxy is very consistent with the value obtained from the simulation snapshot. This is primarily because LAEs are much larger than dwarfs and hence, are not expected to have very strong outflows. This also justifies neglecting the outflow term due to galactic winds in eq. (\ref {md}).

\begin{figure} 
  \center{\includegraphics[scale=0.5]{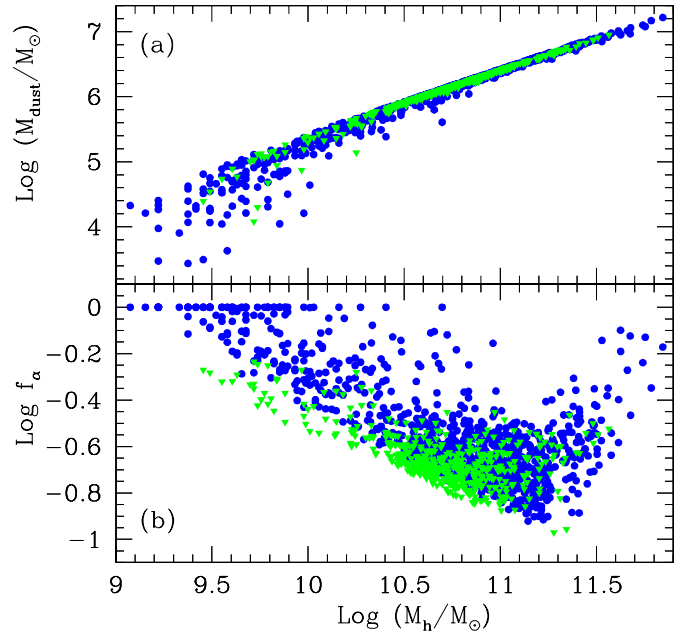}} 
  \caption{(a) Dust mass and (b) $f_\alpha$ as a function of halo mass, $M_h$, for z=5.7 (circles) and z=6.6 (triangles) for the galaxies identified as LAEs at these redshifts using $p=(1.5,0.6)$ for all galaxies at $z\sim(5.7,6.6)$. For details, refer to text in Sec. \ref{dust model}.}
\label{dust} 
\end{figure} 

The destruction timescale, $\tau_{dest}(t)$ is estimated following the results of McKee (1989), Lisenfeld \& Ferrara (1998) as:
\begin{equation}
\tau_{dest} (t) = \frac{M_g(t)}{\gamma \langle \dot M_* \rangle \epsilon M_s {\rm (100\, km \, s^{-1})} } .
\end{equation}
Here, $\epsilon$ is the efficiency of dust destruction in a SN-shocked ISM, for which we adopt value $\sim 0.4$; this is a reasonable estimate between the values of 0.1 and 0.5 found by McKee (1989) and Seab \& Shull (1983) for varying densities of the ISM and magnetic field strengths. $M_s {\rm(100\,  km\, s^{-1}) }$ is the mass accelerated to 100 km ${\rm s^{-1}}$ by the SN blast wave and has a value of $6.8 \times 10^3 \, {\rm M_\odot}$ (Lisenfeld \& Ferrara 1998).

Since we consider a scenario in which the dust amount is regulated only by the SFR, which scales with the halo mass, the dust amounts too scale in the same way, as seen from Panel (a) of Fig. \ref{dust}. However, while the dust mass increases linearly with the halo mass for most of the objects, the scatter at the low mass end arises from feedback regulation of the star formation in these objects. 

Once the final dust mass, $M_{dust}(t_*)$, is calculated for each galaxy in the simulation, we can translate this into an optical depth, $\tau_c$, for continuum photons as
\begin{equation}
\tau_c = \frac{3\Sigma_{d}}{4 a s},
\end{equation}
where $\Sigma_d$ is the dust surface mass density, $a$ and $s$ are the radius and material density of graphite/carbonaceous grains, respectively ($a=0.05 \mu m$, $s = 2.25\, {\rm g\, cm^{-3}}$; Todini \& Ferrara 2001; Nozawa 2003). The dust surface mass density is calculated assuming that grains are concentrated in a radius proportional to the stellar distribution, $r_e$, following the results of Bolton et al. (2008), who have derived fitting formulae to their observations of massive, early type galaxies between $z=0.06-0.36$\footnote{The relation used to obtain the stellar distribution radius, $r_e$, from the observed V band luminosity, $L_V$ is 
\begin{eqnarray}
\log r_e &=& 1.28 \log \sigma_{e2} -0.77 \log I_e -0.09, \\
(M_{dim} / 10^{11} M_\odot) & = & 0.691 (L_V / 10^{11} L_\odot)^{1.29}, \\
\sigma_{e2}& = &(2 G M_{dim} / r_e)^{0.5} \,[100 \, {\rm km/s}], \\
I_e &= & L_V/ (2\pi r_e^2) \,[10^9 L_V/ (L_\odot {\rm kpc}^2)].
\end{eqnarray}
Here, $L_V$ (the luminosity between 4644 and 6296 \AA\, in the observer's frame) is obtained using {\rm STARBURST99} for the SFR, age, metallicity and IMF of the galaxy under consideration. For a galaxy at $z \sim 5.7$ which has $t_*=200$ Myr, $Z=0.2\, {\rm Z_\odot}$, $\dot M_*=1\, {\rm M_\odot yr^{-1}}$, the V band luminosity is 5.05$\times \, 10^{42} \, {\rm erg \, s^{-1}}$ and the corresponding $r_e = 0.35$ kpc. Although we have used the V band luminosity in the observer's frame in this work, using the rest frame V band luminosity does not affect our results in any way, using the same values of the free parameters.}. Though it is not an entirely robust estimate, we extend this result to galaxies at $z~5.7,6.6$ due to the lack of observations about the stellar distribution in high-redshift galaxies.

We calculate $\Sigma_d = M_{dust}(t_*)/(\pi r_d^2)$, where $r_d$ is the effective radius of dust. The best fit to the observed UV LFs requires the dust distribution scale to be quite similar to that of the stellar distribution, such that $r_d =(0.6,1.0) \times r_e$ at $z \sim (5.7,6.6)$.

This optical depth can be easily converted into a value for the escape fraction of continuum photons assuming a slab-like distribution of dust, such that
\begin{displaymath}
f_c = \frac{1-e^{-\tau_c}}{\tau_c}.
\end{displaymath}
We are hence able to determine $f_c$ for each individual galaxy, depending on its intrinsic properties, i.e. the SFR, metallicity, age, gas mass and IMF. 

\begin{figure} 
  \center{\includegraphics[scale=0.45]{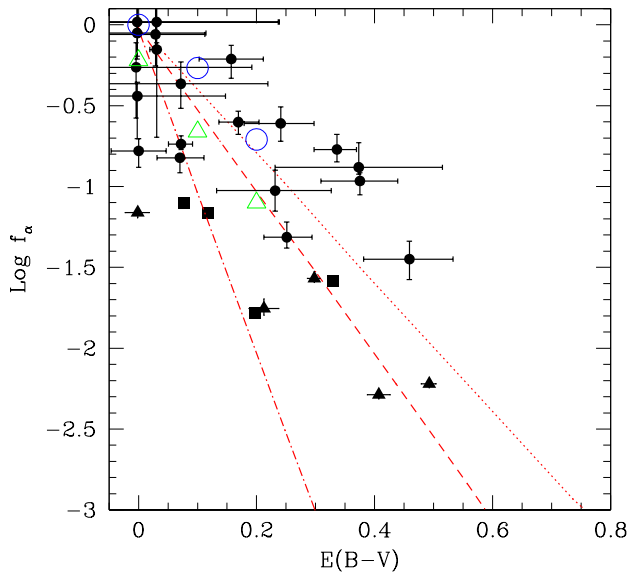}} 
  \caption{The escape fraction of Ly$\alpha$ photons, $f_\alpha$ as a function of the dust color excess, $E(B-V)$. Filled points (circles, triangles, squares) show the observed values for Ly$\alpha$ galaxies from the GALEX LAE sample (Deharveng et al. 2008), IUE local starburst sample (McQuade et al. 1995; Storchi-Bergmann et al. 1995) and Atek et al. (2008) sample respectively. The dashed line is the best fit to all the observations (see Atek et al. 2009), the dot dashed line shows the best fit from Verhamme et al. (2008) obtained from the spectral fitting of $z \sim 3$ LBGs and the dotted line is obtained following the Seaton law (1979). The relations obtained from this work using $p=(1.5,0.6)$ for LAEs are shown with empty circles and triangles for $z \sim (5.7,6.6)$, respectively. Details in Sec. \ref{dust model}.}
\label{fesc} 
\end{figure} 

As $f_c$ is fully determined from the above procedure, we are left with a single free model parameter, $f_\alpha$, to match the Ly$\alpha$ LF obtained from our calculations to the data. 
The relation between $f_\alpha$ and $f_c$ can be written as
\begin{equation}
f_\alpha  = p (A_\lambda, C) f_c,
\end{equation}
where $p$ depends not only on the adopted extinction curve, $A_\lambda$, but also on the differential radiative transfer effects acting on both Ly$\alpha$ and UV continuum photons in an inhomogeneous medium described by a clumping factor $C=\langle n^2 \rangle/\langle n \rangle^2$. While for the former dependence, some pieces of evidence exist that the SN dust we assume here can successfully be used to interpret the observed properties of the most distant quasars (Maiolino et al. 2006) and gamma-ray bursts (Stratta et al. 2007), the way in which inhomogeneities differentially enhance Ly$\alpha$ with respect to UV continuum luminosities through the so-called "Neufeld effect" (Neufeld 1991; Hansen \& Oh 2006; Verhamme et al. 2008; Finkelstein et al. 2009; Kobayashi et al. 2009) is still highly debated. A report of a decreasing trend of $f_\alpha$ with $E(B-V)$ has been recently published (Atek et al. 2009);  our results nicely match that empirical relation, as shown in Fig. \ref{fesc}. We therefore tentatively assume $C=1$ and compute $p$ simply from our SN extinction curve, which corresponds to a value of $R_V=2.06$, obtaining $p (A_\lambda, C=1)=0.8$, independently of the LAE mass or luminosity. Although the best fit to the observed Ly$\alpha$ LF favors values that vary around 0.8 ($p=1.5,0.6$ at $z=5.7,6.6$)\footnote{ We set $p=1$ when $f_\alpha \ge 1$.}, it is unclear if these deviations can be interpreted as a genuine imprint of an evolving clumpy ISM.    

\begin{figure} 
  \center{\includegraphics[scale=0.5]{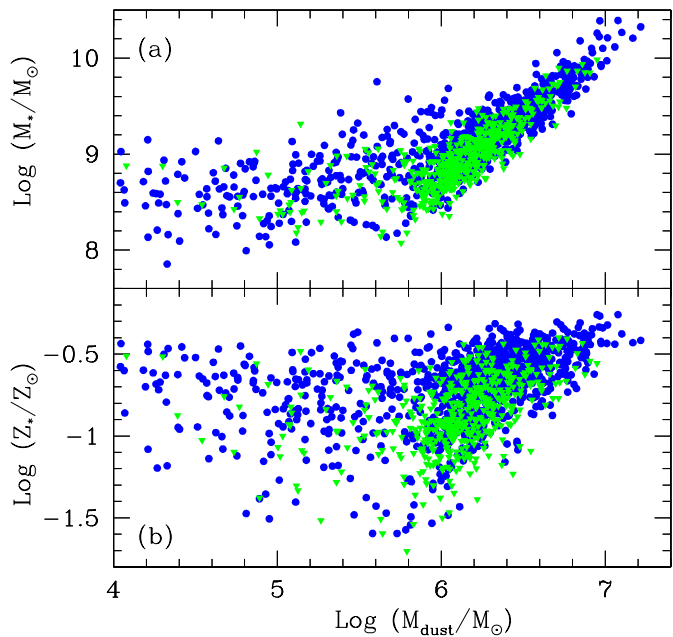}} 
  \caption{Relation between (a) stellar mass, $M_*$, (b) stellar metallicity, $Z_*$ and dust mass, $M_{dust}$, for z=5.7 (circles) and z=6.6 (triangles) for the galaxies identified as LAEs at these redshifts.}
\label{md_met} 
\end{figure} 

We now return to the results obtained using the dust model. For galaxies identified as LAEs at $z \sim 5.7$, the escape fraction of Ly$\alpha$ photons, $f_\alpha$, decreases by a factor of 10, from 1 to about 0.1 as the halo mass increases from $10^{9.0}$ to $10^{11.2}\, {\rm M_\odot}$ as shown in Panel (b) of Fig. \ref{dust}, consistent with the results obtained by Laursen et al. (2009). However, for the few galaxies between $10^{11.2}$ and $10^{11.8}\, {\rm M_\odot}$, $f_\alpha$ increases again. This is because the optical depth, $\tau \propto M_{dust}/r_e^2$, where $r_e$ scales with SFR (see eq. 11). Since the SFR rises steeply with halo mass for $M_h \gsim 10^{11.2}\, {\rm M_\odot}$, the optical depth decreases, even though the total dust mass increases. The decreased optical depth then leads to the larger escape fraction of continuum and hence, Ly$\alpha$ photons seen.

We find that the galaxies with the largest stellar mass are more dust rich, as shown in Panel (a) of Fig. \ref{md_met}; for the bulk of the galaxies, the dust mass, $M_{dust}$ is related to the stellar mass, $M_*$, as $M_{dust} \propto M_*^{0.7} $. However, the low stellar mass end shows a large scatter in the dust content, likely arising from  feedback regulation leading to a scatter in the star formation rates for smaller halos/galaxies. As expected, galaxies with the highest stellar metallicities have the largest dust masses; the dust mass is related to the stellar metallicity, $Z_*$, as $M_{dust} \propto Z_*^{1.7}$ as shown in Panel (b) of Fig. \ref{md_met}. Again, the scatter for the low metallicity galaxies is explained by the corresponding scatter in the star formation rates at this end.

We find that $\langle f_c \rangle \sim (0.23,0.38)$ at $z \sim 5.7, 6.6$ respectively. This is consistent with the values of $(0.22,0.37)$ obtained in Dayal et al. (2009a). The Ly$\alpha$ contribution from the gas allows smaller galaxies to become visible as LAEs: 
the minimum halo mass of LAEs shifts from  $(10^{10.2}) {\rm M_\odot}$ to $(10^{9.0},10^{9.4})$ at $z \sim(5.7,6.6)$. Since smaller galaxies ($M_h \leq 10^{10.2} {\rm M_\odot}$) have lower SFR and dust content, they tend to have a higher $f_c$ value. However, only about 10\% of the galaxies we identify as LAEs in the simulation have $M_h < 10^{10.2} M_\odot$ and hence, they are too few to vary the mean $f_c$.

\section{Observational Implications}
\label{match obs}

\begin{table*} 
\begin{center} 
\caption {For all the LAEs comprising the Ly$\alpha$ LF at the redshifts shown (col. 1), we show the range of halo mass (col. 2), the range of mass weighted ages (col. 3), the range of SFR (col. 4), the range of gas mass (col. 5), the average color excess (col. 6), the average transmission of Ly$\alpha$ photons through the IGM (col. 7), the average intrinsic EW (col. 8), the average EW of the line emerging from the galaxy (col. 9) and the average value of the observed EW (col. 10).}

\begin{tabular}{|c|c|c|c|c|c|c|c|c|c} 
\hline 
$z$ & $M_h$ & $t_*$ & $\dot M_*$ & $M_g$  & $\langle E(B-V) \rangle$ & $\langle T_\alpha \rangle$ & $\langle EW^{int} \rangle$ & $\langle EW^{emer} \rangle$ & $\langle EW \rangle$ \\
$$ & $[M_\odot]$ & $[Myr]$ & $[M_\odot yr^{-1}]$ & $[M_\odot]$ & $$ & $$ & $[{\rm\AA}]$ & $[{\rm\AA}]$ & $[{\rm\AA}]$ \\ 
\hline
$5.7$ & $10^{9.0-11.8}$ & $44.0-326.2$ & $0.8-120$ & $10^{9.0-11.0}$ & $0.14$ & $0.48$ & $106.0$ & $157.2$ & $75.1$  \\
$6.6$ & $10^{9.4-11.6}$ & $23.3-218.1$ & $1.6-46.4$ & $10^{9.1-10.7}$ & $0.09$ & $0.48$ & $117.2$ & $70.3$ & $34.6$ \\ 
\hline
\label{tab1} 
\end{tabular} 
\end{center}
\end{table*} 

\begin{figure} 
  \center{\includegraphics[scale=0.45]{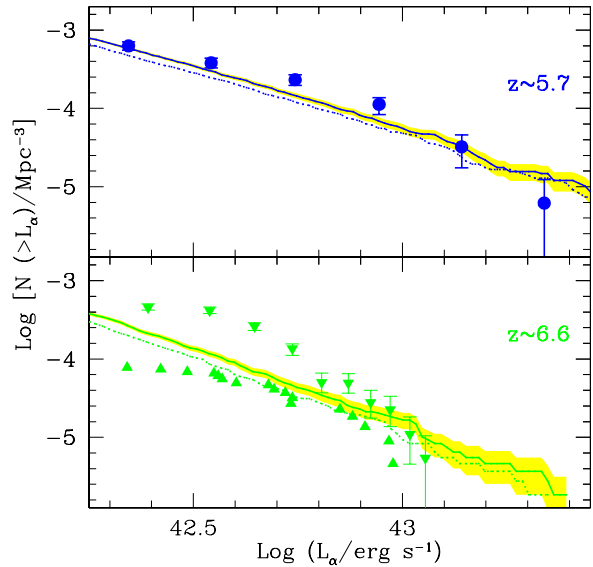}} 
  \caption{Cumulative Ly$\alpha$ LF for the ERM. The panels are for $z\sim 5.7$ and $6.6$. Points represent the data at two different redshifts:  $z\sim 5.7$ (Shimasaku et al. 2006) (circles), $z \sim 6.6$ (Kashikawa et al. 2006) with downward (upward) triangles showing the upper (lower) limits. Solid (dashed) Lines refer to model predictions at $z\sim 5.7,6.6$ for the parameter values in text including (excluding) the contribution from cooling of collisionally excited \HI in the ISM. Shaded regions in both panels show poissonian errors. }
\label{lya} 
\end{figure}

\begin{figure} 
  \center{\includegraphics[scale=0.45]{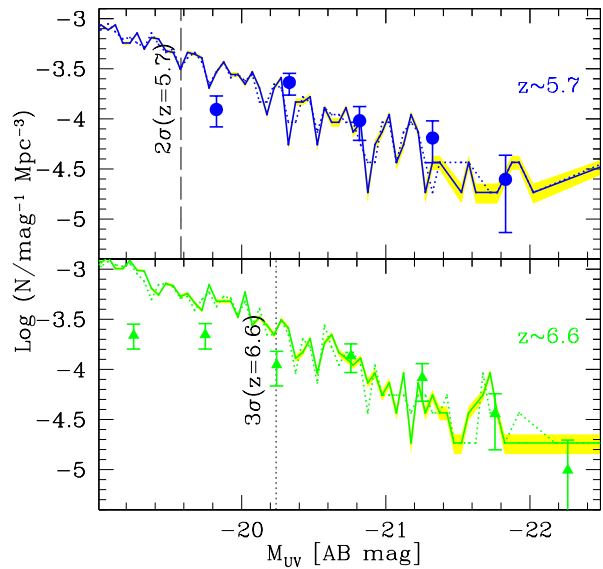}} 
  \caption{UV LAE LF for the ERM. Points represent the data at two different redshifts: $z \sim 5.7$ (Shimasaku et al. 2006) (circles), $z \sim 6.6$ (Kashikawa et al. 2006) (triangles). Solid (dashed) lines refer to model predictions including (excluding) the contribution from cooling \HI in the ISM at the redshifts (from top to bottom): $z \sim 5.7,6.6$, for the parameter values in text. The vertical dashed (dotted) lines represent the observational 2$\sigma$ (3$\sigma$) limiting magnitude for $z = 5.7$ ($z =6.6$). The shaded regions in both panels show the poissonian errors.} 
\label{uv} 
\end{figure}  

\begin{figure} 
\center{\includegraphics[scale=0.45]{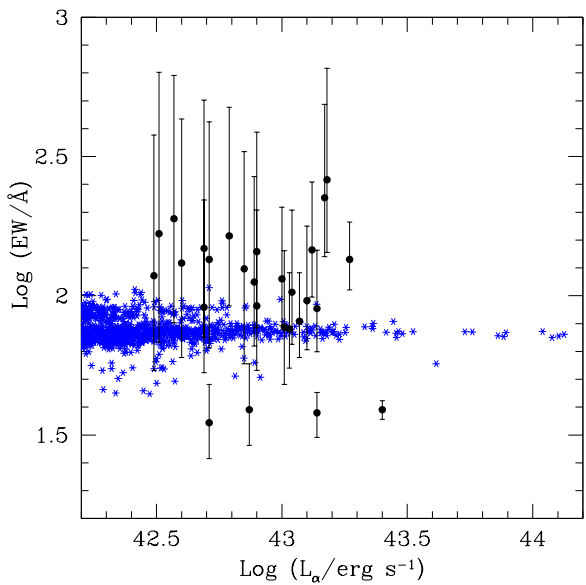}} 
\caption{Observed EWs from Shimasaku et al. (2006) (circles) and model values of the observed EWs (astrexes) as a function of the observed Ly$\alpha$ luminosity.}
\label{ewfig57} 
\end{figure}  

As mentioned before, cooling of collisionally excited \HI can produce a significant amount of Ly$\alpha$ luminosity in the ISM. However, since galaxies become more compact and denser towards high redshift, this leads to a corresponding increase in $L_\alpha^g$. This can be clearly seen from the cumulative Ly$\alpha$ LF (Fig. \ref{lya}) where ignoring the gas contribution leads to a slightly larger under estimation of the LF at $z \sim 6.6$ compared to $z\sim 5.7$. 

As mentioned in Sec. \ref{lyag}, cooling of collisionally excited \HI does not contribute much to the continuum luminosity, with the average contribution to the total continuum luminosity being $\leq 10$ \% at either of the redshifts considered. Adding $L_c^g$, therefore,  produces a negligible effect on the UV LF, as seen clearly from Fig. \ref{uv}, where the UV LFs, both including and excluding the gas contribution overlap. 

The free parameters in this work are the the radius of the dust distribution, $r_d$ relative to the stellar distribution radius, $r_e$ and the escape fraction of Ly$\alpha$ photons relative to continuum photons. Once that these two parameters are obtained by calibrating our results to the UV and Ly$\alpha$ LFs, we have no more free parameters left. Using the values of $f_\alpha/f_c$, $T_\alpha$ and the intrinsic EW for each galaxy we identify as a LAE, we obtain the emergent and observed EWs at $z \sim 5.7$. Due to the cooling of collisionally excited \HI, the total Ly$\alpha$ luminosity increases by a factor of about 1.12 while the continuum almost remains unchanged. This leads to intrinsic EWs in the range 88.8-188.8 \AA \, for the galaxies we identify as LAEs. The average value of the intrinsic EW at $z \sim 5.7$ also increases to 106.0 \AA\, compared to 95 \AA\, that we found in Dayal et al. (2009a), not including gas emission.

The EWs emerging from the galaxy at $z \sim 5.7$ are further enhanced and have a mean value of 157.2 \AA\, because of the larger escape fraction of Ly$\alpha$ photons with respect to continuum photons ($f_\alpha/f_c=1.5$). However, since the Ly$\alpha$ luminosity is attenuated by transmission through the IGM, with $\langle T_\alpha \rangle \sim 0.48$, the observed EWs are finally reduced to the range 44.4-106.9 \AA; with an average value of 75.1 \AA. 

We show the EWs as a function of the observed Ly$\alpha$ luminosity in Fig. \ref{ewfig57}. Though the numerical data is concentrated at ${\rm Log}(EW/{\rm \AA}) \sim 1.9$, it shows some scatter, mostly induced by the different physical properties of galaxies as age and metallicity. However, we are unable to reproduce some of the largest and smallest EW values. This could be due to some of the assumptions made in the modelling of $L_\alpha^g$ including:  (a) we have calculated $L_\alpha^g$ for a homogeneous ISM. This assumption seems to represent a fairly good description of the low-$z$ data discussed in Fig. 4. However, we cannot exclude the possible presence of gas inhomogeneities which would increase the collisional excitation rates and hence increase $L_\alpha^g$. Since we are investigating a large cosmological volume, resolving the small scale structure of the ISM is beyond the possibilities of the present work, and our results in this sense should be considered as a lower limit to the true emissivity of the gas. (b) $L_\alpha^g$ depends on the radius of gas distribution and the simple relation we have assumed might be different for these high redshift galaxies and (c) we have assumed a very simple scaling between $f_\alpha$ and 
$f_c$. In reality, this should depend on the geometry and distribution of dust within the ISM and this would induce scatter in the EW distribution. 

\section{Summary} 
\label{summary}

Coupling SPH simulations with a Ly$\alpha$ production/transmission model, in Dayal et al. (2009a) we studied the evolution in the properties of Ly$\alpha$ emitters at high redshifts. In this paper, we further extend that work by including two important ingredients, based on the intrinsic properties of each galaxy: (a) Ly$\alpha$ and continuum luminosities produced by cooling of collisionally excited \HI with electrons in the ISM and (b) a dust model to calculate the optical depth and hence, the escape fraction of continuum photons (and color excess) from each galaxy. 

We find that the Ly$\alpha$ luminosity from cooling ISM gas ($L_\alpha^g$) constitutes a non-negligible addition to the value from stars ($L_\alpha^*$); the average value of $L_\alpha^g / L_\alpha^* \sim (0.16,0.18)$ at $z \sim (5.7,6.6)$. However, the continuum from stellar luminosity ($L_c^*$) always dominates over that from cooling gas ($L_c^g$); the average value of $L_c^g / L_c^* \sim (0.01,0.02)$ at $z \sim (5.7,6.6)$. The contribution of the cooling of \HI to both the Ly$\alpha$ and continuum luminosities increases with redshift because of the increasing \HI gas density, which leads to a larger excitation and hence, cooling rate. Since the SFR rises steeply with increasing halo mass, $L_\alpha^*$ and $L_c^*$ also increase faster as compared to $L_\alpha^g$ and $L_c^g$ and hence dominate in massive galaxies.

The dust budget of a given LAE is determined by several physical processes (dust formation, destruction and astration) mainly regulated by the evolution of massive stars. Hence
the dust mass increases with increasing halo mass, following the behavior of the star formation rate. Galaxies with the largest stellar mass, $M_*$ have also the largest dust content, with a scaling relation given by  
$M_{dust} \propto M_*^{0.7}$. The same trend of increasing dust content is also observed in terms of the stellar metallicity, $M_{dust} \propto Z_*^{1.7}$.

We calculate the escape fraction of continuum photons using a slab-like dust distribution model and graphite/carbonaceous grains (radius $a=0.05 \mu m$, density $\rho= 2.25 {\rm g cm^{-3}}$). We then assume a mass-independent relation to go from $f_c$ to $f_\alpha$ such that $f_\alpha=(1.5,0.6) f_c$ at $z \sim (5.7,6.6)$ respectively. The increase in the dust mass with increasing halo mass leads to a decrease in the escape fraction of Ly$\alpha$ photons ($f_\alpha$) from the galaxy, with the value dropping from $\sim 1$ to $0.1$ as the halo mass increases from $10^{9.0}$ to $10^{11.2}\, {\rm M_\odot}$ at $z \sim 5.7$. 


Using a SN dust extinction curve (Todini \& Ferrara 2001), $f_\alpha = 0.8 f_c$. This means that the ISM dust is quite homogeneous for $f_\alpha \leq 0.8 f_c$; a situation we find at $z \sim 6.6$. However, no extinction curve can produce the value of $f_\alpha > f_c$, which hints at the fact that the dust could be clumped at $z \sim 5.7$ as proposed by Neufeld (1991) and Hansen \& Oh (2006).

Using two free parameters to translate $f_c$ to $f_\alpha$ and $r_e$ to $r_d$, we are able to reproduce the UV and Ly$\alpha$ LFs reasonably at both the redshifts considered. With no more free parameters, we calculate the intrinsic, emergent and observed EWs from the galaxies we identify as LAEs. Adding Ly$\alpha$ from the cooling \HI, the average intrinsic EW at $z\sim 5.7$ increases by a factor of about 1.12 and has a mean vale of $106.0$ \AA. The emergent EW is further enhanced to an average value of about 157.2 \AA\, at $z \sim 5.7$ since $f_\alpha/f_c=1.5$ at this redshift. However, attenuation of the Ly$\alpha$ luminosity \HI in the IGM reduces the mean observed EW to 75.1 \AA.

The observed EWs (Shimasaku et al. 2006) show a large scatter when plotted against the observed Ly$\alpha$ luminosity. By virtue of using a Ly$\alpha$ production/transmission and dust model that depends on the intrinsic properties of each galaxy, we also obtain a scatter in the same plot.

We conclude by discussing some unsolved points still left in our calculations. Though we are able to reproduce the Ly$\alpha$ LF at $z\sim 5.7$ to a reasonable extent, we are still unable to capture the shape of the slope completely. Further, though we get a scatter in the EWs, we are still unable to produce the largest and smallest EWs. These could be due 
to the  simplifying assumptions adopted in our modelling. These include: (a) a simple scaling relation between the gas distribution scale and the virial radius for each galaxy. However, a spread in this relation would vary the $L_\alpha^g$ calculated; (b) assuming a clumpy ISM would increase $L_\alpha^g$ as compared to the value we get assuming a homogeneous distribution; (c) a simple scaling between $f_\alpha$ and $f_c$. Note, however, that a relation dependent on the individual ISM conditions of each galaxy would induce a scatter in the EW distribution; (d) inflows/outflows, could vary the EWs and probably also leave an imprint on the Ly$\alpha$ LF and (d) a detailed radiative transfer model of reionization which could affect our calculation of $T_\alpha$.
All these issues will be investigated in future works with the aim of building a complete and coherent model for LAEs.

\section*{Acknowledgments} 
We thank the anonymous referee for valuable comments. We thank S. Borgani and L. Tornatore for useful discussions and collaboration. PD would like to thank SNS Pisa, where this work was partly carried out, for hospitality.


\newpage 
\label{lastpage} 

\begin{thebibliography}{99} 

\bibitem[\protect\citeauthoryear{Anderson et al.}{2000}]{a} Anderson H., Balance C.P., Badnell N.R., Summers H.P., 2000, J. Phys. B, 33, 1255

\bibitem[\protect\citeauthoryear{Atek et al.}{2008}]{ha} Atek H., Kunth D., Hayes M., Ostlin G., Mas-Hesse J.M., 2008, A\&A, 488, 491

\bibitem[\protect\citeauthoryear{Atek et al.}{2009}]{ha} Atek H., Kunth D., Schaerer D., Hayes M., Deharveng J.M., Ostlin G., Mas-Hesse J.M., 2009, arXiv Preprint: 0906.5349

\bibitem[\protect\citeauthoryear{Bianchi \& Schneider}{2007}]{bs} Bianchi S., Schneider R., 2007, MNRAS, 378, 973

\bibitem[\protect\citeauthoryear{Bolton et al.}{2008}]{bol} Bolton A.S., Treu T., Koopmans L.V.E., Gavazzi R., Moustakas L.A., Burles S., Schlegel D.J., Wayth R., 2008, ApJ, 684, 248

\bibitem[\protect\citeauthoryear{Cen}{1992}]{cen} Cen R., 1992, ApJS, 78, 341

\bibitem[\protect\citeauthoryear{Dawson et. al.}{2007}]{daw} Dawson S., Rhoads J. E., Malhotra S., Stern D., Wang J., Dey A., Spinrad H., Jannuzi B. T., 2007, ApJ, 671, 1227 

\bibitem[\protect\citeauthoryear{Dayal et al.}{2008}]{me1} Dayal P., Ferrara A. \& Gallerani S., 2008, MNRAS, 389, 1683

\bibitem[\protect\citeauthoryear{Dayal et al.}{2009a}]{me2} Dayal P., Ferrara A., Saro A., Salvaterra R., Borgani S., Tornatore L., 2009a, arXiv Preprint: 0907.0337

\bibitem[\protect\citeauthoryear{Deharveng et al.}{2008}]{dv} Deharveng J.-M. et al., 2008, ApJ, 680, 1072

\bibitem[\protect\citeauthoryear{Dijkstra, Lidz \& Wyithe}{2007a}]{dlw} Dijkstra M., Lidz A., Wyithe J. S. B., 2007a, MNRAS, 377, 1175

\bibitem[\protect\citeauthoryear{Dijkstra, Wyithe \& Haiman}{2007b}]{dwh} Dijkstra M., Wyithe J. S. B., Haiman Z.,  2007b, MNRAS, 379, 253

\bibitem[\protect\citeauthoryear{Dijkstra}{2009}]{d9} Dijkstra M., 2009, ApJ, 690, 82 

\bibitem[\protect\citeauthoryear{Dijkstra \& Loeb}{2009}]{b} Dijkstra M., Loeb A., 2009, arXiv preprint: 0902.2999

\bibitem[\protect\citeauthoryear{Dwek, Galliano \& Jones}{2007}]{dgj} Dwek E., Galliano F. \& Jones A.P., 2007, Preprint: arXiv:0711.1170

\bibitem[\protect\citeauthoryear{Ferrara, Pettini \& Shchekinov}{2000}]{r2} Ferrara A., Pettini M. \& Shchekinov Y., 2000, MNRAS, 319, 539

\bibitem[\protect\citeauthoryear{Finkelstein et al.}{2009}]{f9} Finkelstein S.L., Rhoads J.E., Malhotra S., Grogin N., 2009, ApJ, 691, 465

\bibitem[\protect\citeauthoryear{Gallerani et al.}{2007}]{b6} Gallerani S., Ferrara A., Fan X., Choudhury T.R., 2007, arXiv Preprint: 0706.1053 

\bibitem[\protect\citeauthoryear{Gnedin et al.}{2008}]{hmr} Gnedin N.Y., Kravtsov A.V., Chen H.W., 2008, ApJ, 672, 765

\bibitem[\protect\citeauthoryear{Gronwall et al.}{2007}]{g7} Gronwall C. et al., 2007, ApJ, 667, 79

\bibitem[\protect\citeauthoryear{Haardt \& Madau}{1996}]{ham} Haardt F., Madau P., 1996, ApJ, 461, 20

\bibitem[\protect\citeauthoryear{Hansen \& Oh}{2006}]{ho6} Hansen M., Oh S.P., 2006, MNRAS, 367, 979

\bibitem[\protect\citeauthoryear{Kashikawa et al.}{2006}]{b10} Kashikawa N. et al., 2006, ApJ, 648, 7 

\bibitem[\protect\citeauthoryear{Kobayashi et al.}{2007}]{kob} Kobayashi M.A.R., Totani T., Nagashima M., 2007, ApJ, 670, 919 

\bibitem[\protect\citeauthoryear{Kobayashi et al.}{2009}]{kob} Kobayashi M.A.R., Totani T., Nagashima M., 2009, arXiv preprint: 0902.2882

\bibitem[\protect\citeauthoryear{Komatsu et al.}{2009}]{kom} Komatsu E. et al., 2009, ApJS, 180, 330

\bibitem[\protect\citeauthoryear{Lai et. al.}{2007}]{lai} Lai K., Huang J., Fazio G., Cowie L.L., Hu E.M., Kakazu Y., 2007, ApJ, 655, 704

\bibitem[\protect\citeauthoryear{Laursen et al.}{2009}]{pl} Laursen P., Sommer-Larsen J, Andersen A.C., arXiv Preprint: 0907.2698

\bibitem[\protect\citeauthoryear{Leitherer et al}{199}]{le} Leitherer C. et al., 1999, ApJS, 123, 3

\bibitem[\protect\citeauthoryear{Lisenfeld \& Ferrara}{1998}]{r4} Lisenfeld U. \& Ferrara A., 1998, ApJ, 496, 145

\bibitem[\protect\citeauthoryear{Maiolino et al.}{2006}]{rm} Maiolino R. et al., 2006, MmSAI, 77, 643

\bibitem[\protect\citeauthoryear{Mao et al.}{2007}]{m7} Mao J., Lapi A., Granato G.L., De Zotti G., Danese L., 2007, ApJ, 667, 655 

\bibitem[\protect\citeauthoryear{McKee}{1989}]{r5} McKee C.F., 1989, Proc IAU symp. 135, Interstellar Dust. Kluwer, Dordrecht, p.431

\bibitem[\protect\citeauthoryear{McQuade et al.}{1995}]{mcqu} McQuade K., Calzetti D., Kinney A.L., 1995, ApJS, 97, 331 

\bibitem[\protect\citeauthoryear{McQuinn et al.}{2007}]{mcq} McQuinn M., Hernquist L., Zaldarriaga M., Dutta S., 2007, MNRAS, 381, 75 

\bibitem[\protect\citeauthoryear{Murayama et al.}{2007}]{m7} Murayama T. et al., 2007, ApJS, 172, 523 

\bibitem[\protect\citeauthoryear{Mori, Umemura \& Ferrara}{2004}]{mmf} Mori M., Umemura M., Ferrara A., 2004, ApJ, 613, 97

\bibitem[\protect\citeauthoryear{Mori \& Umemura}{2007}]{mu} Mori M., Umemura M., 2007, Ap\&SS, 311, 111

\bibitem[\protect\citeauthoryear{Nagamine et al.}{2008}]{ng} Nagamine K., Ouchi M., Springel V., Hernquist L., 2008, arXiv preprint: 0802.0228 

\bibitem[\protect\citeauthoryear{Neufeld}{1991}]{neu} Neufeld D.A., 1991, ApJ, 370, 85

\bibitem[\protect\citeauthoryear{Nozawa et al.}{2003}]{n3} Nozawa T., Kozasa T., Umeda H., Maeda K., Nomoto K., 2003, ApJ, 598, 785

\bibitem[\protect\citeauthoryear{Nozawa et al.}{2007}]{n7} Nozawa T., Kozasa T., Habe A., Dwek E., Umeda H., Tominaga N., Maeda K., Nomoto K., 2007, ApJ, 666, 955

\bibitem[\protect\citeauthoryear{Osterbrock}{1989}]{os} Osterbrock D.E., 1989, Astrophysics of Gaseous Nebulae and Active Galactic Nuclei. University Science books, Sausalito, CA 

\bibitem[\protect\citeauthoryear{Ouchi}{2008}]{b} Ouchi M. et al., 2008, ApJS, 176, 301

\bibitem[\protect\citeauthoryear{Padovani \& Matteucci}{1993}]{pad}Padovani P., Matteucci F., 1993, ApJ, 416, 26


\bibitem[\protect\citeauthoryear{Salpeter}{1955}]{sal} Salpeter E. E., 1955, ApJ, 121, 161

\bibitem[\protect\citeauthoryear{Saro et al.}{2006}]{saro} Saro A., Borgani S., Tornatore L., Dolag K., Murante G., Biviano A., Calura F., Charlot S., 2006, MNRAS, 373, 397

\bibitem[\protect\citeauthoryear{Santos}{2004}]{san} Santos M. R., 2004, MNRAS, 349, 1137

\bibitem[\protect\citeauthoryear{Schaerer}{2003}]{s3} Schaerer D., 2003, A\&A, 397, 527

\bibitem[\protect\citeauthoryear{Seab \& Shull}{1983}]{ss} Seab C.G., Shull J.M., 1983, ApJ, 275, 652

\bibitem[\protect\citeauthoryear{Seaton}{1979}]{st} Seaton M.J., 1979, MNRAS, 187, 73

\bibitem[\protect\citeauthoryear{Shimasaku et al.}{2006}]{b17} Shimasaku K. et al., 2006, PASJ, 58, 313

\bibitem[\protect\citeauthoryear{Spitzer \& Greenstein}{1951}]{sg5}  Spitzer L.J., Greenstein J.L., 1951, ApJ, 114, 407

\bibitem[\protect\citeauthoryear{Springel et. al.}{2001}]{sp1} Springel V., Yoshida N., White S.D.M., 2001, New Astronomy, Vol 6, Issue 2, pg.79 

\bibitem[\protect\citeauthoryear{Springel \& Hernquist}{2003}]{mcq} Springel V., Hernquist L., 2003, MNARS, 339, 289

\bibitem[\protect\citeauthoryear{Springel}{2005}]{spa} Springel V., 2005, MNRAS, 364, 1105

\bibitem[\protect\citeauthoryear{Storchi-Bergmann et al.}{1995}]{sb} Storchi-Bergmann T., Kinney A.L., Challis P., 1995, ApJS, 98, 103

\bibitem[\protect\citeauthoryear{Stratta et al.}{2007}]{st} Stratta G., Maiolino R., Fiore F., D'Elia V., 2007, ApJ, 661, 9

\bibitem[\protect\citeauthoryear{Sutherland \& Dopita}{1993}]{mcq} Sutherland R.S., Dopita M.A., 1993, 88, 253

\bibitem[\protect\citeauthoryear{Todini \& Ferarra}{2001}]{tf} Todini P. \& Ferarra A., 2001, MNRAS, 325, 726.

\bibitem[\protect\citeauthoryear{Tornatore et. al.}{2007}]{mcq} Tornatore L., Borgani S., Dolag K., Matteucci F., 2007, MNRAS, 382, 1050  

\bibitem[\protect\citeauthoryear{Valiante et al.}{2009}]{rosa} Valiante R., Schneider R., Bianchi S., Andersen A.C., 2009, arXiv Preprint: 0905.1691

\bibitem[\protect\citeauthoryear{van den Hoek et. al.}{1997}]{van} van den Hoek L.B., Groenewegen M.A.T., 1997, A\&A Supp., 123, 305

\bibitem[\protect\citeauthoryear{Verhamme et al.}{2008}]{av} Verhamme A., Schaerer D., Atek H., Tapken C., 2008, A\&A, 491, 89

\bibitem[\protect\citeauthoryear{Woosley \& Weaver}{1995}]{woo} Woosley S.E., Weaver T.A., 1995, ApJS, 101, 181

 
 
\end{thebibliography}
\end{document}